\documentclass[galaxies,review,accept,oneauthor,pdftex]{Definitions/mdpi}

\firstpage{1} 
\pubvolume{1}
\issuenum{1}
\articlenumber{0}
\pubyear{2021}
\copyrightyear{2021}
\externaleditor{Academic Editor: Margo Aller, Jose L. Gómez and Eric Perlman}
\datereceived{13 April 2021} 
\dateaccepted{1 August 2021} 
\datepublished{} 
\hreflink{https://doi.org/} 

\Title{Message in a Bottle: Unveiling the Magneto-Ionic Complexity of AGNs through the Stokes QU-Fitting Technique}

\TitleCitation{Message in a Bottle: Unveiling the Magneto-Ionic Complexity of AGNs through the Stokes QU-Fitting Technique}

\Author{Alice Pasetto  $^{1,2}$\orcidA{}}

\AuthorNames{Alice Pasetto}

\AuthorCitation{Pasetto, A.}

\address{%
$^{1}$ \quad Instituto de Radioastronom\'ia y Astrof\'isica (IRyA-UNAM), 3-72 (Xangari), 8701, Morelia, M\'exico. a.pasetto@irya.unam.mx\\
$^{2}$ \quad Consejo Nacional de Ciencia y Tecnolog\'ia (CONACyT), Av. Insurgentes Sur 1582, Col. Cr\'edito Constructor,
Alcald\'ia Benito Ju\'arez, C.P. 03940, Ciudad de M\'exico.}

\abstract{Here, I overview one of the available techniques for the analysis of broad-band spectropolarimetric data, the Stokes QU-fitting. Since broad-band receivers have been installed at most radio facilities, the collection of radio data, both the total intensity and the linear polarization, is revealing interesting features in their spectra. The polarized light, and therefore its properties, i.e. the fractional polarization \textit{p} and the polarized angle $\chi$, are now finally well sampled in wide wavelength ranges. The new complex behaviors revealed by the data can be studied using the Stokes QU-fitting, which consists of modeling the Stokes parameters \textit{Q} and \textit{U} using wavelength-dependent analytical models, available in the literature. This technique provides a very good diagnostic of the nature and structure of the magnetized plasma, with the possibility to identify complex structures, internal or external, of the source of study. A summary of the available and most used models describing the polarization behavior, is presented. Moreover, some of the most significant observational works which use this technique are also summarized.}


\begin{document}


\section{Introduction}
An important  astrophysical topic that is still a matter of investigation is the understanding of the physical properties of magnetised plasma. It is still unclear under which conditions, and how, the magnetic field influences the formation and evolution of astrophysical objects, what its role is in fundamental processes such as galaxy formation, particle acceleration or confinement of the Active Galactic Nuclei (AGN) jets.  It is indisputable that the magnetic field (B-field) is present everywhere in the Universe and at different physical scales. It is possible to investigate the B-field with different observational methods, at very different wavelengths, directly through polarized observations in the electromagnetic spectrum, or indirectly through, for example, magnetic Kelvin--Helmholtz (KH) instabilities (e.g.,~\cite{Vikhlinin2001}).
In the radio regime, an important radiative process, which provides information about the magnetic field, is the synchrotron radiation. Measuring the wavelength dependence of the polarized light from synchrotron sources can reveal valuable and interesting constraints on the nature of the objects observed, such as insights into the magneto--ionic medium internal to the source of investigation or about their surroundings. The quantities extracted directly from the observations are the Stokes parameters, \textit{I}, \textit{Q}, \textit{U} and \textit{V}, where \textit{I} represents the total intensity, \textit{Q} and \textit{U} represent the linear polarized signal and \textit{V} the circular polarization. In this review, only the properties of the linear polarization will be discussed; therefore, the Stokes \textit{V} is not taken into consideration. From the Stokes parameters \textit{I}, \textit{Q} and \textit{U}, one can estimate the fractional polarization (\textit{p}) and the polarized angle ($\chi$). The first gives an idea of the ratio of the turbulent over ordered magnetic field strength projected in the plane of the sky, while the latter provides information about the orientation of the magnetic field projected on the sky. Moreover, the synchrotron polarized light can experience two effects: Faraday rotation and Faraday depolarization. Both these effects can alter the polarization properties (\textit{p} and $\chi$) of the radiation. The study of both phenomena allows us to obtain information about the strength, the direction and the complexity (ordered or turbulent) of the magnetic field, identified by the parameters of Faraday Depth (FD - $\phi$) or the Rotation Measure (RM) dispersion ($\sigma_{RM}$). Measurements and analysis of these quantities can help to define a 3-dimensional (3D) view of the magneto--ionic medium in astrophysical sources. 

In general radio polarimetric studies have been considered very challenging. Since the polarized light is only a small fraction of the total emission from a source, it is necessary to observe with high sensitivity in order to perform a proper polarization analysis. 
Moreover, the collection of sparse and narrow-band polarized data makes difficult to fully describe the polarization behavior of the subject of study. Nevertheless, important works focused on identifying the magneto--ionic structure of radio sources were performed in the last decades using different radio telescopes, different frequencies and different angular resolutions (e.g., \cite{Cotton1984,Dreher1987,Attridge1999,Berkhuijsen2003,Zavala2003}). Many spectral studies of source-integrated, radio band linear polarization from the Michigan program, starting in the mid 1960s, and from the Brandeis group using the Very Long Baseline Array (VLBA) data, gave important results \cite[see][and references therein]{Wardle2021}. Simultaneously, important theoretical works have also been developed in order, for example, to explain the observed brightness and polarization asymmetries between the jets and counter-jets for some Fanaroff--Riley Class I (FR I) radio galaxies. The asymmetries are explained with a model in which decelerating backflows, containing a toroidal magnetic field, are covering the central jets' spine~\cite{Laing2012,Laing2014}. Moreover, besides using the traditional determination and interpretation of the RM (i.e., as the linear fit of the polarized angle in a range of sparse wavelength squared) to extract information about the magnetic structure of astrophysical objects, an alternative technique was developed. It combines the study of the spectral index (called spectral tomography \cite{KatzStone1997}) with the polarized emission information. This technique also produced important results, such as the detection of the jet spine and sheath components in wide-angle tails' radio sources (see \cite{KatzStone1999, Hardcastle1999}). 

Thanks to recent upgrades performed at some radio facilities, for example, at the Jansky Very Large Array (Jansky VLA) and at the Australia Telescope Compact Array (ATCA), and the new incoming generation of ultrasensitive radio telescopes, such as the Low-Frequency Array (LOFAR \cite{vanHaarlem2013}), the Murchison Widefield Array (MWA \cite{Tingay2013}), the Australian Square Kilometer Array Pathfinder (ASKAP \cite{McConnell2016}) or the future Square Kilometer Array (SKA), which are (or will be) equipped with broad-band receivers, radio polarization observations have become a much more captivating and fashionable research area. Thanks to this new capability, our ability to interpret polarised sources has dramatically improved. Now, we are able to collect well sampled polarized data and therefore to follow the behavior of the polarization properties in a broader frequency domain.  It is possible to use the Faraday effect to create a tomographic reconstruction of magnetized structures along the line of sight (LOS) and, therefore, to determine the distribution, and analyze the properties, of the magneto--ionic material along the LOS internal or external of an astronomical object. This process is called Faraday tomography (see the Galaxies MDPI special issue of the  workshop---The Power of Faraday Tomography---for more information \url{https://www.mdpi.com/journal/galaxies/special_issues/FaradayTomography}, accessed on November 2018),
which allows to add information on the origin and structure of magnetic fields. As a consequence, new computational techniques for the analysis of the wide-band spectropolarimetric data have been developed. The two most commonly used techniques are the Faraday RM synthesis \cite{Brentjens2005} and the Stokes QU-fitting \cite{Farnsworth2011,OSullivan2012}. The first is a Fourier transform operation on the polarized radiation in the wavelength domain;  it assumes the target as a sum of emitters at different Faraday depths. Because of its nature, the RM synthesis can be considered a non-parametric technique.  There are several works in the literature using this technique  (e.g., see recent works by \cite{Loi2019,Krause2020,Stuardi2020}), a detailed explanation of which is out of the scope of this work. In contrast, the QU-fitting technique, which is the focus of this review,  is a model fitting method  (usually computationally more expensive but more powerful than RM Synthesis  \cite{Sun2015}), which consists of fitting both the Stokes \textit{Q} and \textit{U} parameters, wavelength ($\lambda$) dependent, using analytical models available in the literature. This technique provides an excellent diagnostic of the nature and structure of magnetized plasma. It has the advantage that physically motivated models can be fit directly to the polarization data, therefore yielding to physical constraints on the magneto--ionic structure of the sources. These models assume the presence of a regular and/or a turbulent magnetic field (e.g., \cite{Burn1966, Tribble1991, Sokoloff1998}). For this reason, the Stokes QU-fitting needs some assumptions on the nature of the medium. The main parameters in the models are the intrinsic fractional polarization (\textit{${p_0}$}), the intrinsic polarization angle ($\chi_{0}$), the Faraday Depth ($\phi$) and the RM dispersion ($\sigma_{RM}$). Generally, the fit model attempts to describe the complex physics that the radio object is experiencing. For that reason, sometimes it is necessary to use multiple Faraday components (i.e., a combination of models) to better describe the polarized~behavior. 

This work is structured as follows: in Section \ref{MODELS} I summarize the most commonly used models for the Stokes QU-fitting technique, in Section \ref{RECENTOBS}, I review some of the recent and most significant observational works on spectropolarimetry with a special focus on AGN jets, and in Section \ref{CONCLUSION}, a summary and conclusions are given.


\section{Polarization Models}
\label{MODELS}

Analytical models that attempt to describe polarization behavior  have been available since the 1960s. They explore different scenarios in which uniform and/or turbulent magnetic fields are involved, according to the results of radio observations (for details see the works by \cite{Burn1966, Tribble1991, Sokoloff1998, Rossetti2008}). For an exhaustive review on the polarization of extragalactic radio sources, please see \cite{Saikia1988}.

To interpret the information of the polarized light of a source which passes through a medium, it is necessary to define the observables. The synchrotron radiation from a radio source is intrinsically partially polarized and its polarization signal can be represented by its Stokes parameters, \textit{I}, \textit{Q}, \textit{U} and \textit{V}. These parameters have the advantage that they are directly measured from the radio telescope and describe the polarization characteristics of a source. Following \cite{Sokoloff1998}, these observables can be combined to describe the linear polarized signal as a complex number:
\begin{equation}
P=Q+iU=pIe^{2i\chi},
\end{equation} 
where $\chi$ is the observed  electric E-vector polarization angle. Using the notation in \cite{Farnsworth2011} we define the fractional Stokes parameters, \textit{q = Q/I} and \textit{u = U/I}, 
so that the fractional (linear) polarization $\textit{p}$ is defined as:
\begin{equation}
p=\sqrt{q^2+u^2},
\end{equation}
and the polarized angle is obtained from:
\begin{equation}
\chi[rad]=\frac{1}{2} {\rm arctan}(\frac{u}{q}).
\end{equation}

Faraday rotation and, consequently, the Faraday depolarization, are two phenomena that can complicate the behavior of the linearly polarized optically thin synchrotron emission. In fact, when one, or both, of these events take place, the polarized angle and the fractional polarization can be altered. The polarized wave can be decomposed into two opposite-handed circularly polarized components. When the radiation passes through a magneto--ionic medium, these opposite-handed waves experience different phase velocities within the material.  This results in a rotation of the polarized plane of the linearly polarized wave, called the Faraday rotation. The general expression of the Faraday depth is given by the equation:
\begin{equation}
\phi [rad/m^{2}]=8.1\times10^5\int_Ln_eB_\parallel dL ,
\end{equation}
where $n_{e}$ [cm$^{-3}$] is the electron density in the plasma, $B_\parallel$ [Gauss] is the magnetic field along the line of sight and L is the path length [pc]. Narrow-band multi-wavelength polarization observations have shown that the polarization angle changes linearly as a function of $\lambda^2$ so that:
\begin{equation}
\Delta\chi [rad]=\chi(\lambda)-\chi(0)=\phi\lambda^2.
\end{equation}

From these equations, it becomes clear that a detailed study of the polarization properties across the spectrum can provide physical information about the source and its surroundings. In the presence of a uniform magnetic field in the medium, the Faraday depth is identical to the polarized emission-weighted mean value, called the Rotation Measure (RM), which is therefore determined by fitting the linear equation $\chi (\lambda^2) =\chi_0+RM\lambda^2$. However, this equation is true only when one restricts the RM fitting to regions of the $\lambda^2$ space, where the fractional polarization, p is constant (e.g.,  \cite{Simard-Normandin1981}). In this scenario, the complex representation of the polarized wave in the presence of a single Faraday RM screen, which only rotates the polarized plane of the light, leaving the fractional polarization constant, becomes:
\begin{equation}
P=p_0e^{2i(\chi_0 + RM\lambda^2)},
\label{eq1}
\end{equation}
where $p_0$ and $\chi_0$ are the intrinsic fractional polarization and the intrinsic polarization angle, respectively. This is the only situation where d$\chi$/d$\lambda^2$ and p($\lambda^2$) are constant, that is, only a single uniform foreground Faraday screen is present. Real scenarios are much more complicated and more sophisticated modeling is required. Since the spectropolarimetric data obtained from broad-band receivers have become available, the behavior of the main polarization parameters, $\textit{p}$ and $\chi$, have been observed to behave in a complex way. The general behavior of $\chi$ is very far from a simple linear variation with $\lambda^2$, and the $\textit{p}$ is usually reduced at longer wavelengths producing a Faraday depolarization \mbox{(e.g.,~\cite{Law2011, OSullivan2012,Anderson2015, Anderson2016, Pasetto2016, Pasetto2018})}. Thus, it is necessary to study these properties across the spectrum in order to properly obtain the physical parameters of the sources \cite{Farnsworth2011}. The first exhaustive work that examines the polarized emission for a subsample of sources taken from the NRAO VLA Sky Survey (NVSS) catalogue was done by \cite{Farnes2014}. Although they collected the available narrow-band polarized data, they could already fit the depolarization behaviors, using simple descriptions of the more complex depolarization models.  

Depolarization can be caused internally within the source, when the synchrotron emitting and the Faraday rotating regions are spatially coincident, or externally to the source, when the Faraday rotation region is located between the source and the observer. Moreover, both scenarios can be described, considering the presence of either a uniform magnetic field, a turbulent magnetic field or a combination of both.  For an exhaustive description of the different scenarios, see \cite{Sokoloff1998}. Depolarization effects should therefore be added to the complex linear polarization Equation \eqref{eq1}.  When the synchrotron radiation passes through an external magneto--ionic medium, that is, a region devoid of relativistic electrons, containing a turbulent magnetic field, the plane of polarization undergoes a random walk which leads to depolarization as long as many turbulent cells are within the beam area. For a regular magnetic field, depolarization can occur when variations in the strength or direction of the field within the observing beam, exist. This depolarization is called External Faraday Dispersion/Beam depolarization (EFD/BD) and it is represented by the equation
\begin{equation}
P=p_0e^{-2\sigma^2_{RM}\lambda^4}e^{2i(\chi_0+\phi\lambda^2)},
\label{eq2}
\end{equation}
where $\sigma_{RM}$ is the Faraday dispersion of the random field within the volume traced by the telescope beam \cite{Farnsworth2011} and $\phi$ is the Faraday depth across the source on the sky and it assumes the form of RM when considering the presence of a regular magnetic field.

When the synchrotron-emitting and the Faraday rotating regions are embedded together, the depolarization is called internal. The main equations which describe this depolarization scenario are: (1) internal Faraday dispersion (IFD) and (2) the differential Faraday rotation (DFR) \cite{Burn1966, Sokoloff1998}.  When the synchrotron-emitting and the Faraday-rotating regions contain a mixture of a turbulent and regular magnetic fields, the equation which takes into account both scenarios is the following:
\begin{equation}
P=p_0e^{2i\chi_0}\left(\frac{1-e^{-S}}{S}\right),
\label{eq3}
\end{equation}
where S = 2$\sigma^2_{RM}\lambda^4$--2i$\phi\lambda^2$.  This equation includes the DFR  as the imaginary term and the IFD as the real term. Here, a random walk of the plane of polarization through the region, occurs. In this equation, $\sigma_{RM}$ is the Faraday dispersion of the internal random field and again $\phi$ is the Faraday depth through the region. When the turbulent magnetic field component is neglected ($\sigma_{RM}$ = 0), the emitting and rotating regions are only co-spatial in the presence of a regular magnetic field. In this situation, the DFR takes place and the complex degree of polarization is given by: 
\begin{equation}
P=p_0e^{2i(\chi_0+\frac{1}{2}\phi\lambda^2)}\left(\frac{{\rm sin}( \phi\lambda^2)}{(\phi\lambda^2)}\right).
\label{eq4}
\end{equation}

In this case, the radiation coming from the most distant part of the region with respect to the observer suffers a different amount of Faraday rotation (in this case $\phi$ is a parameter whose value is to be estimated) with respect to the radiation coming from the nearest part of that region. Therefore, this equation also embraces the scenario in which a smooth change in the Faraday depth occurs within a source. Here, the Faraday depth $\phi$ assumes the form of a simple RM, which produces a gradient of RM ($\Delta$RM)  across the telescope beam \citep{Berkhuijsen1990,Sokoloff1998}. Equation \eqref{eq4} becomes: 
\begin{equation}
P=p_0e^{2i(\chi_0+RM\lambda^2)}\left(\frac{{\rm sin}( \Delta RM\lambda^2)}{(\Delta RM\lambda^2)}\right).
\label{eqBurn}
\end{equation}

However, the parameter $\Delta$RM can describe both internal and external Faraday depolarization effects; for example, a linear gradient in RM across the emission region, or internal Faraday rotation in a uniform field \citep[][]{Sokoloff1998,Schnitzeler2015}.
The internal case is of great interest since it should represent the simplest scenario occurring within a radio jet where a smooth change of the magnetic field configuration across the jet is expected \citep[][]{Laing1981,Blandford1993}. 

More generally, the polarization dependencies given by Equations \eqref{eq2}--\eqref{eqBurn} could not be uniquely associated with the given distributions of emitting and thermal plasma. For example, the oscillatory behavior of the fractional polarization is not uniquely associated with the presence of the mixing of thermal and non-thermal plasma. In fact, it can be produced by a combination of several narrow Faraday screens of the Faraday dispersion function (FDF, equation \eqref{eq1}). This scenario describes a regular magnetic field but it does not specify between the foreground or the mixed model. The same occurs when talking about the external Faraday rotation. It can be represented by the FDF, located in the foreground of a mixed plasma, but represented with a Gaussian-like shape. Equations \eqref{eq2}--\eqref{eqBurn}, correspond to FDF that can be produced by specific and more complex configurations of thermal and non-thermal plasma. Therefore, it is necessary to support the analysis of the polarized data with additional information, for example, the angular resolution, or from the analysis of the continuum synchrotron emission to draw a more complete picture of the magneto--ionic medium of the object of study.

\begin{figure}[H]
\widefigure
\includegraphics[width=0.8\textwidth]{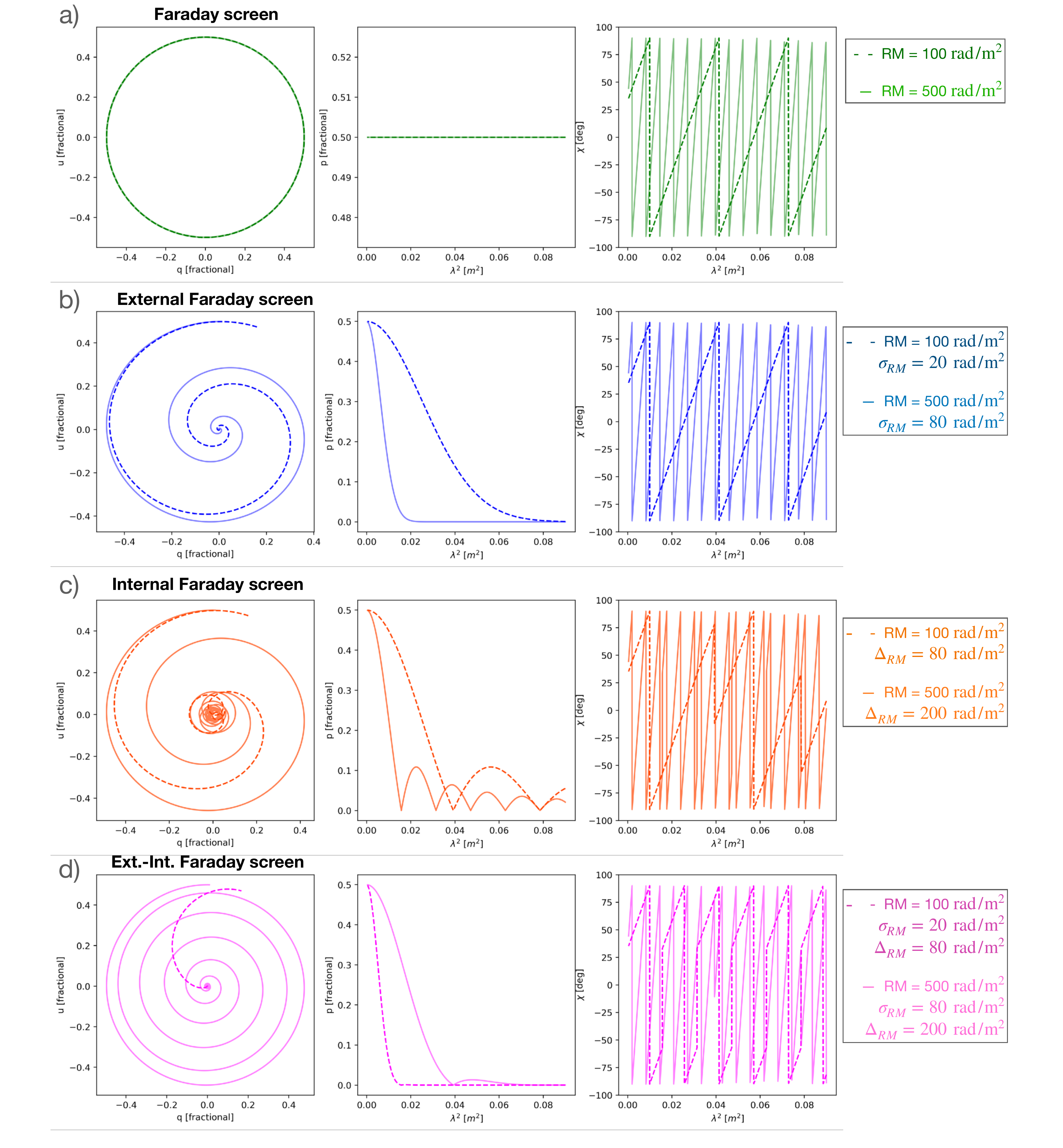}
\caption{Representation of de/polarization models: (\textbf{a})  Equation \eqref{eq1}, Faraday rotation (only one Faraday component does not depolarize the emission); (\textbf{b}) Equation \eqref{eq2}, External Faraday depolarization (considering turbulent magnetic field); (\textbf{c})~Equation~\eqref{eq4}, Internal Faraday depolarization (considering regular magnetic field, only);  (\textbf{d}) Equation \eqref{eq5}, Mixture of Internal and External Faraday depolarization. Comparison between the behaviors of low and large RM values and low and large $\sigma_{RM}$ and $\Delta_{RM}$ are represented by dashed and straight lines, respectively. The used values of RM, $\sigma_{RM}$ and $\Delta_{RM}$, are reported in the labels on the right side of the plots. }
\label{fig:AllModels}

\end{figure}

To investigate the complex behavior of the polarized signal, it is therefore necessary to fit simultaneously the wide band Stokes \textit{Q} and \textit{U} spectra as proposed by \cite{Farnsworth2011, OSullivan2012}.  To capture a broad range of possible Faraday rotation behaviors, we use a general equation which attempts to describe all the above mentioned depolarization models (describing the effects of both random and uniform magnetic fields), which is the following complex polarization equation:
\begin{equation}
P= p_{0}e^{2i(\chi_{0}+\frac{1}{2}\phi\lambda^2)} sinc(\phi\lambda^2) e^{-2\sigma^2_{RM}\lambda^4},
\label{eq5}
\end{equation}
 where \textit{p$_{0}$} is the intrinsic fractional polarization, $\chi_{0}$ is the intrinsic polarization angle, and $\phi$ and $\sigma_{RMj}$ describe the variation of the Faraday depth in a regular and turbulent magnetic field, respectively.
This equation uses the simplest possible parameterization of the Faraday depolarization from uniform and random fields. It is also implicitly assumed that the polarized emission is coming from optically thin regions with similar spectral indices.  However, the above equation is too simple for explaining the complex scenarios revealed by real astrophysical data. Instead, multiple emitting and/or rotating Faraday components unresolved within the telescope beam seem to better represent the reality in most cases  (see, e.g.,  \cite{OSullivan2012,Anderson2016,Pasetto2018}). Therefore, to model more complicated behaviors, extra RM components need to be used. A multiple Faraday components model is simply constructed as P = P$_{1}$ + P$_{2}$ +...+ P$_{n}$ \cite{OSullivan2012}. A representation of the models mentioned above is shown in Figure \ref{fig:AllModels}. 

The operational procedure to perform the fitting is explained below. Firstly we need to collect wide-band polarized data, which, after a standard cleaning process, result in a sample of Stokes \textit{I}, \textit{Q} and \textit{U} images at each available frequency (within the wide-band frequency range). The subsequent step depends on the scientific case to address. The polarized Stokes parameters can be extracted at each frequency considering either the object of study in its totality (in the case of both resolved or unresolved radio sources) or by selecting specific areas. Lastly, the parameters are then used for the model fit, applying Equation \eqref{eq5}. The results of the fitting are the intrinsic polarized properties of the magnetized layer/s ($p_{0j}$, $\chi_{0j}$, $\phi_j$ and $\sigma_{RMj}$ ) which describe a scenario where the Faraday screen/s is/are either internal or external to the source of study, depending on the best resulting fit model, or a combination of models. The common statistic used to determine the goodness of the fitting is the standard ${\chi^2}$ test. In general, the ${\chi^2}$ value decreases with an increasing number of initial parameters. However, choosing a model with too many parameters can result in an overfitting. Therefore, two other statistical methods are usually used in combination with the ${\chi^2}$ test: the Akaike Information Criterion (AIC) and the Bayesian Information Criterion (BIC). AIC estimates the quality of each model, relative to each of the other models considered in the study, while BIC is a criterion for model selection among a sample of models.

\begin{figure}[H]
\widefigure
\includegraphics[width=0.8\textwidth]{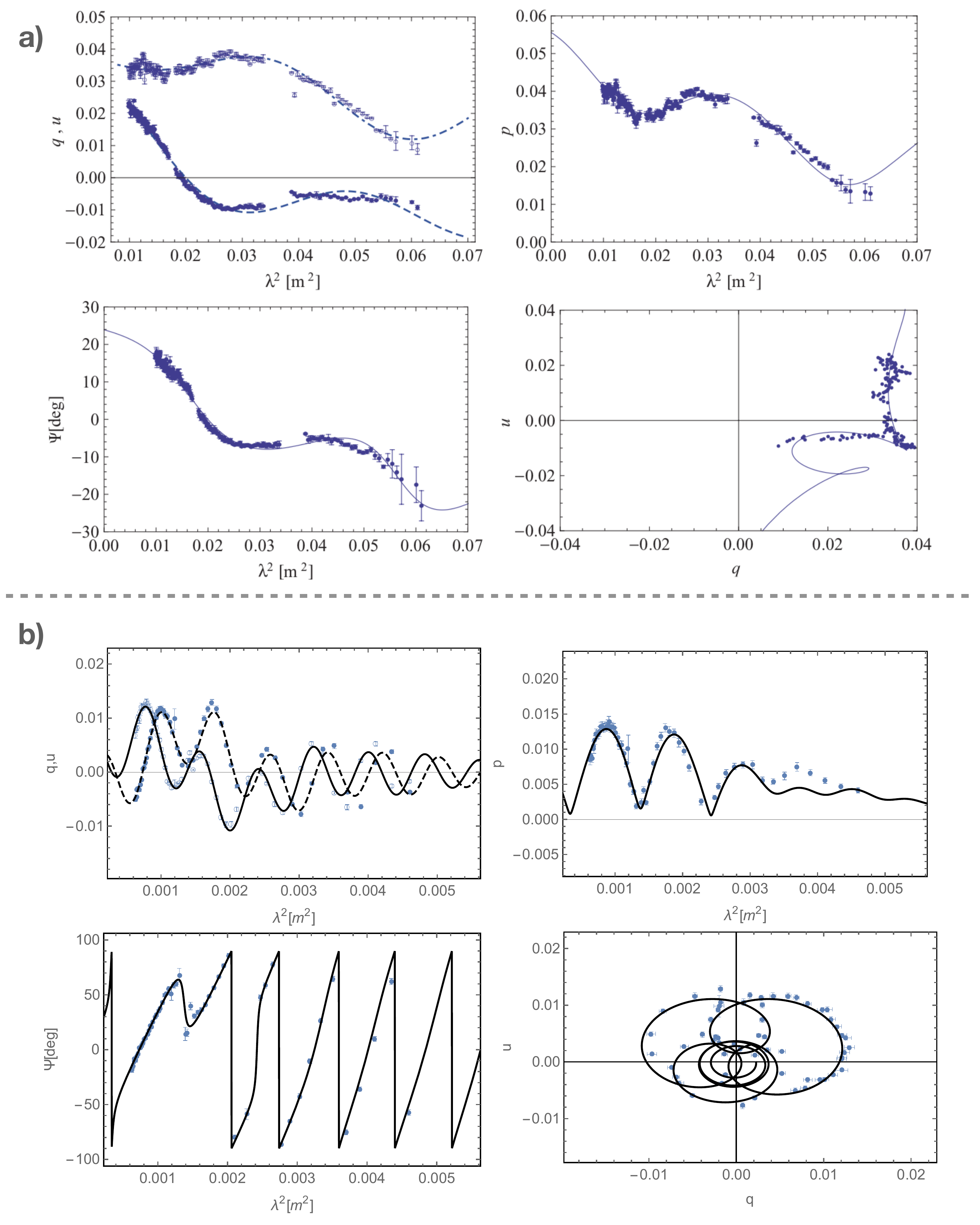}
\caption{Depolarization Stokes QU-fitting on the sources: {(\textbf{a})} PKSB1039-47 and {(\textbf{b})} 0958+3224. The four panels represent: the Stokes parameters \textit{q}, \textit{u} (\textbf{top left}), the fractional polarization \textit{p} (\textbf{top right}), the polarized angle $\chi$ [deg] (\textbf{bottom left}) all versus wavelength squared $\lambda^2$[m$^2$] and the Stokes \textit{u} versus Stokes \textit{q} (\textbf{bottom right}). Straight lines are the resulting model fit. Both sources have been fitted with three internal Faraday screens (Equation  \eqref{eqBurn}). Images adapted from \cite{OSullivan2012,Pasetto2018}.}
\label{fig:shanealice}

\end{figure}

\begin{figure}[H]
\widefigure
\includegraphics[width=0.76\textwidth]{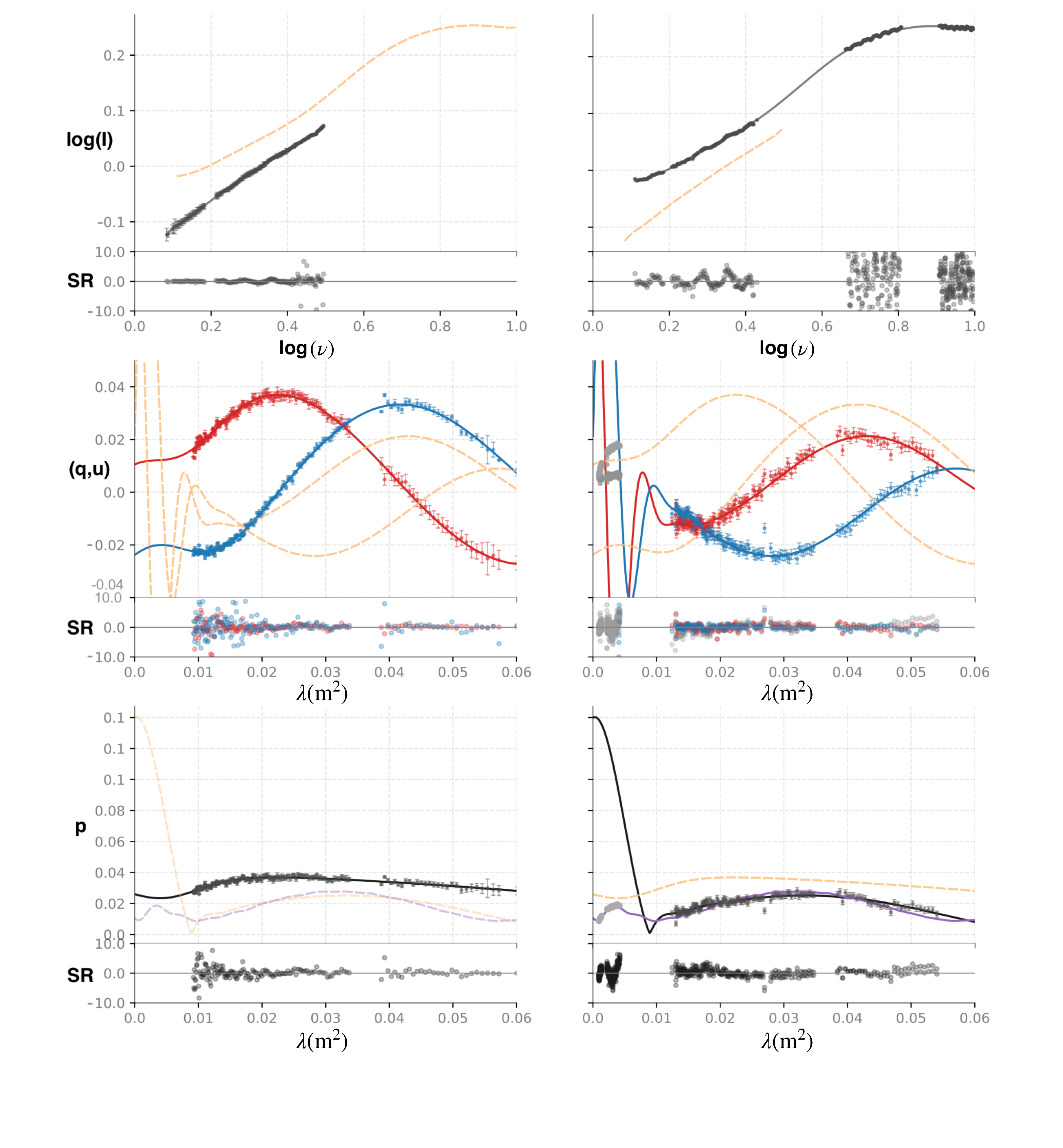}
\caption{Stokes \textit{I} data (black points, top row of panels), Stokes \textit{q} and \textit{u} data (red and blue points, middle row) and fractional polarisation data (\textit{p}, black points, last row) for PKS B0454-810 in the 2012 and 2017 epochs (left and right column of panels, respectively). To facilitate comparison between epochs, the best-fit model for both the 2012 and 2017 epochs are mirrored on the opposite epoch axes as dashed orange lines for each of the Stokes \textit{I}, (\textit{q},\textit{u}), and \textit{p} plots.  Straight lines (black, red and blue) are the best model fit. The sub-axes on each panel are the standardised residuals (measured in standard deviations) for the difference between the data and the best-fit model. Image adapted from \cite{Anderson2019}.}
\label{fig:craig}

\end{figure}

\clearpage
\end{paracol}
\nointerlineskip
\vspace{-8pt}
\begin{paracol}{2}
\switchcolumn



\section{Observational Broad-Band Spectropolarimetry Works}
\label{RECENTOBS}

\subsection{Unresolved Sources}

Although early analyses performing model fit on sparse narrow-band fractional polarization and polarization angle (using the traditional linear fit of $\chi$ versus $\lambda^2$) radio data, are present in the literature, (e.g., \cite{Laing1987,Laing2008,Rossetti2008, Guidetti2010}), the first call of attention about the importance of fitting the Stokes \textit{Q} and \textit{U} in broad $\lambda^2$ domain was made by \cite{Farnsworth2011}.  They presented a polarization analysis of 585 compact radio sources observed with the WSRT at 350 MHz. On some of the sources, they performed different analyses: the traditional $\chi(\lambda^2)$ linear fit, the Stokes QU-fitting and the RM Synthesis techniques. They fit the sources using two simple depolarization scenarios: a foreground screen and two interfering RM components. They found that, although a single representative RM may be found using all the model fitting techniques, the Faraday structure of the sources is not always adequately represented. They stressed the importance of considering both the degree and the angle of polarization (or equivalently, \textit{q} and \textit{u}) over a range of $\lambda^2$ as wide as possible, in order to obtain a more global picture of the polarization behavior. They favour QU-fitting as the technique which allows a more accurate description of the Faraday structure/s.

A subsequent work opened the door to the broad-band spectropolarimetry study on AGN jets: O'Sullivan et al. \cite{OSullivan2012} 
presented a detailed study of the Faraday depth structure of four bright (>1 Jy), strongly polarized, point-like radio-loud quasars. Using the ATCA telescope at frequencies from 1 to 3 GHz (2 GHz of instantaneous bandwidth) they fit various Faraday rotation models to the data. From the results of the broad-band fitting it was claimed that the polarized structures arose from the compact inner regions of the radio sources themselves  (either from their intrinsic emissions or from their local environments) and not from polarized emissions from galactic or intergalactic foreground regions. This result has important implications for using background extragalactic radio sources to probe the Galactic and intergalactic magneto--ionic media. In fact, narrow-band observations can produce erroneous results whenever a source shows multiple interfering Faraday components. Moreover, this result  suggests that RM time variability in extragalactic point sources may be due to the evolution of polarized components, which reveal different parts of a magneto--ionic medium surrounding the radio jet. 

The first VLA broad-band (1--12 GHz) spectropolarimetric study of a sample of point-like AGN was performed by \cite{Pasetto2018}. The sample (14 sources) consisted of very compact sources (with linear  sizes smaller than $\sim$5 kpc) that are unpolarized at 1.4 GHz in the NVSS survey. The sources  are highly polarized at high frequency and therefore characterized by large Faraday RM values.  Highly reliable complexity (with very high signal to noise -S/N- detection of $\sim$1000 for the continuum data and $\sim$100 for the polarized signal), has been detected. The sources show several total intensity synchrotron and Faraday components (up to three Faraday screens) in both their continuum and polarized spectra. It is worth mentioning that more components (synchrotron components or Faraday screens) are more likely to be fitted with large S/N data.  Depolarization modeling was performed on the polarized data and a better knowledge on the AGN jet environment for that sample, was achieved. The work concluded that for the majority of the targets, the very large RMs and strong depolarization were due to turbulent magnetic fields local to the sources. The synchrotron radiation from the jet passes through several clumpy or highly turbulent regions which cover the central jet. Similar conclusions were obtained by \cite{Anderson2016}, by analyzing a sample of unresolved sources. They also detected complex behaviors in almost every source in the sample mostly due to Faraday dispersive medium intrinsic to the source themselves. Examples of the depolarization model fitting on point-like sources are shown in Figure~\ref{fig:shanealice}.

O'Sullivan et al. \cite{OSullivan2017} presented the results of a broad-band (1 to 3 GHz), spectropolarimetry study from 100 extragalactic point like radio sources with the ATCA, selected to be highly linearly polarized at 1.4 GHz. The science case behind the project was to investigate whether some correlation between the broad-band polarized behavior of the radio source and their host galaxy accretion states existed. They divided the sample into: ``radiative-mode'' (objects with strong, high-ionization, optical emission lines powered  by accretion onto the central super-massive black holes at rates in excess of $\sim$1 per cent of the Eddington limit) and ``jet-mode'' (objects with weak or non-existent optical emission lines, possibly due to radiatively inefficient accretion) AGN. They fit multiple Faraday components, up to three distinct polarized emission regions, to the majority of their sources and they found no significant difference between the Faraday rotation or Faraday depolarization properties of jet-mode and radiative-mode AGN. However, a statistically significant difference in the intrinsic degree of polarization between the two types has been found: the jet-mode sources showed more intrinsically ordered magnetic field structures than the radiative-mode sources. A preferred perpendicular orientation of the intrinsic magnetic field structure to the axes of FRI jets, is considered a possible explanation for the high intrinsic degree of polarization in the ``jet-mode'' sources. The authors also found that almost all FRs are hosted by low-excitation/jet-mode AGN. No clear magnetic field orientation preference was found for the radiative-mode sources. This may be related to the inner jet regions of FRI radio galaxies where the magnetic field is expected to have a high degree of~order.

\end{paracol}
\nointerlineskip
\begin{paracol}{2}
\switchcolumn


Finally, it is also worth mentioning the work by \cite{Anderson2019}, in which they could detect the evolution of polarized structures on a sample of blazars through the analysis of broad-band spectropolarimetic polarization emission variability. The sources were selected from the~catalogues in \cite{OSullivan2012,Anderson2016}, which are dominated by sub-kpc-scale emission structures. They analyzed temporal changes between two epochs separated by $\sim$5 years using ATCA data. Polarized complexity with the presence of several Faraday components and variability in their polarization emission, were detected. They concluded that the observed spectral changes most likely originate because of the evolution in the parsec to decaparsec scale structure of the blazar jets themselves. Figure~\ref{fig:craig} shows an example of the polarized spectral changes between the two epochs.

\end{paracol}
\nointerlineskip
\begin{paracol}{2}
\switchcolumn

All the above works show how the QU-fitting technique is essential to investigating the true Faraday depth structure; this technique is necessary for probing the magnetized radio source environment and to finally spectrally resolve the polarized components of unresolved radio sources.

\subsection{Resolved Sources}

Recently, the community is putting lot of efforts in studying the magneto--ionic media surrounding AGN jets of nearby, well resolved radio objects. 
Pioneer works on understanding the magneto--ionic media around resolved radio jets are those, for example, in \cite{Owen1990, Laing2008}, in which the authors observed M87 and 3C31 with the VLA at four and six discrete narrow-band frequencies, respectively. The objective was to explore the spatial scale and the origin of the RM in these sources. They concluded that, in both cases, the main reason for Faraday rotation and depolarization behaviors are due to foreground material.  Now, with the available broad-band data, the community can perform more sophisticated QU-fitting modeling, unveiling much more complex structures within or around the sources of study.

Banfield et al. \cite{Banfield2019} showed broad-band radio polarization observations of NGC 612 taken with the ATCA from 1 to 3 GHz. They examined the linear polarization, Faraday rotation, and depolarization distributions resulting from the interaction between the lobes of the radio galaxy NGC 612 and the circumgalactic gas. They found that the polarized properties are tracing a behavior, likely due to an external Faraday rotation characterized by a turbulent magnetic field in the intra-group medium. Evidence of a correlation between the Faraday structure and the radio morphology was found. In particular, the increase in the RM and the depolarization at the hotspot is most likely explained by the compression of the ambient gas due to the bow-shock region on to the lobe.

The radio lobes of the radio galaxy Fornax A have been analyzed by \cite{Anderson2018FornaxA}. A detailed study of both the continuum and the polarized radiations, were performed using the ATCA at frequencies from 1 to 3 GHz. They could map with great detail the magnetic field of the radio lobes.  The Stokes QU-fitting revealed an oscillatory depolarization behavior, therefore leaving  open two scenarios where the inhomogeneities in the magneto--ionic structure of the thermal material, or a co-located suffusion of synchrotron-emitting plasma, are responsible for the large Faraday dispersions and/or differential Faraday depths which are generating the associated depolarization. The toy model proposed suggests the presence of advection instabilities in the lobes where material from the interstellar medium (ISM) are mixing with the lobes' synchrotron radiation, producing filamentary structures where the complex polarized behaviors are detected.

Zooming into a nearby relativistic radio jet and to study its spectropolarimetric properties, is the objective of the work by \cite{Pasetto2021}. Broad-band VLA data, from 4 to 18 GHz, of the kpc jet of M87 is presented and the  Stokes QU-fitting analysis is performed to investigate the nature of its magnetic field.  M87 has a powerful jet which emits synchrotron radiation in a broad frequency domain (i.e., up to optical wavelengths) and therefore has been extensively studied in the past. There have been observations in the radio and optical regimes, and its magneto--ionic medium analyzed (e.g., \cite{Owen1980,Owen1989,Owen1990,Capetti1997,Heinz1997,Perlman1999,Zavala2003,Avachat2016}). Recently, important studies on the Faraday rotation have been performed at different radio frequencies and different radio linear scales (e.g., \cite{Chen2011, Algaba2016, Park2019,Kravchenko2020}). These works claim the detected Faraday rotation to be associated with an external Faraday screen, located in the vicinity of the jet. In particular, the claimed external layer is attributed either to an un-collimated wind launched from hot accretion flows \cite{Park2019,Kravchenko2020} or to the sheath of the jet with a non-negligible additional contribution from the lobe (part of which should be located in front of knot C \cite{Algaba2016}). However, the resolution and the sensitivity of these sparse narrow-band radio data did not allow the authors to resolve the polarized emission across the M87 jet nor to properly study the polarization properties across the frequency domain. Nevertheless, the Stokes QU-fitting has been performed on sparse narrow-band VLBA data to test a possible internal Faraday rotation \cite{Park2019}, reaching no significant conclusion on this matter.

 With the new broad-band, well sampled, VLA data, Reference \cite{Pasetto2021} wanted to test the specific model of the internal Faraday rotation which would suggest a gradual change in the direction of the magnetic field, hence the presence of a helical magnetic field. For a helical configuration, it is expected to observe gradients of the fractional polarization and the Faraday depth perpendicular to the jet flow as the line-of-sight magnetic field changes its direction (see for details: \cite{Laing1981, Blandford1993, Lyutikov2005}). Thanks to the high sensitivity data, depolarization modeling (by using the Stokes QU-fitting) at each pixel has been performed. Maps of the polarization properties, that is, intrinsic fractional polarization and intrinsic polarized angle, have been obtained from the fitting. For the first time, a highly detailed map, that is, with a high significance polarization signal, of the magnetic field configuration (with spatial resolution of tens of pc, see Figure~\ref{fig:BfieldZooms}), has been obtained. At the position of the core, the magnetic field configuration shows a change in its direction recalling a convex angle (see zoom of the VLA core in Figure~\ref{fig:BfieldZooms}). Along the jet, the B-field is well ordered at least up to knot A. However, it also shows regions, mainly between the knots, where the B-field lines behave like a spiral (between knot D and E and between F and I) or they open like a funnel (between know E and F). Moreover, changes by 90 degrees at the position of knot A (see zoom of knot A in Figure~\ref{fig:BfieldZooms}), where a shock front occurs, are well visible. Drastic changes at knot C mark a more chaotic configuration at the termination of the jet (see zoom of knot C in Figure~\ref{fig:BfieldZooms}).  An RM map of the entire kpc M87 jet showing variations of RM along the jet (see Figure \ref{fig:RMmap}) has been obtained. Internal depolarization model  fits have been performed integrating the polarized signal  over some specific areas, especially at the position of knots A, B and C. The result of the depolarization fit are shown in \mbox{Figure~\ref{fig:DepolABC}} and the  resulting parameters are listed in Table~\ref{tab:DepolABC}. The Stokes QU-fitting highlights that  at least three internal Faraday screens are needed at each of these positions to explain the polarization behavior. The internal depolarization behavior  and the presence of variations of the RM values, are features which suggest the presence of a helical magnetic field configuration in a jet \cite{Laing1981,Blandford1993,Lyutikov2005}. The Stokes QU-fitting allows us to better constrain the configuration and the nature of the B-field. An exhaustive analysis on this matter will soon be published in a forthcoming paper.

 Polarized emission coming from nearby spiral galaxies with the objective to understand the nature and the evolution of their large-scale ordered magnetic fields has also been extensively studied. To pursue this objective, several fitting techniques have been performed in the past. The three-dimensional model of the magnetic field represented in terms of a Fourier series has been applied to study the spiral galaxy M51. This method, using multi-frequency, narrow-band polarization observations, allows the identification of the existence of a magneto--ionic halo and to show that the magnetic fields in the disk and the halo have different configurations \cite{Berkhuijsen1997}. With broad-band data, it is now possible to also use the Stokes QU-fitting for the analysis of the polarization emission, to enrich the knowledge of and to map the galaxies' magneto--ionic media. For example, the system of the Antennae galaxies has been studied using both the techniques of Rotation Measure Synthesis and Faraday depolarization analysis. Detailed analysis of the intrinsic polarization properties (i.e., the intrinsic fractional polarization and the intrinsic polarization angle, both results of the depolarization modeling) revealed the presence of turbulent magnetic field within the merging bodies but also a large-scale regular field structure in the outskirts, most likely generated by the stretching of the galactic disc during the interaction \cite{Basu2017}. 
Finally, the Stokes QU-fitting has been used to unveil and analyze the Faraday structures of a lensing disk galaxy. Its coherent $\mu$G magnetic field has been detected with similar strength and geometry to local volume galaxies. This has strong implications for the formation and evolution of the magnetic fields, in particular the observed coherent magnetic field property is compatible with a mean-field dynamo origin \cite{Mao2017}.

The Stokes QU-fitting technique has been used for several other topics, for example to study the ambient medium in galaxy clusters and to study the magnetic field in cosmic web filaments (e.g., the recent work by \cite{Ozawa2015,Mizener2020,DiGennaro2021}).

\begin{figure}[H]
\includegraphics[width=0.8\linewidth]{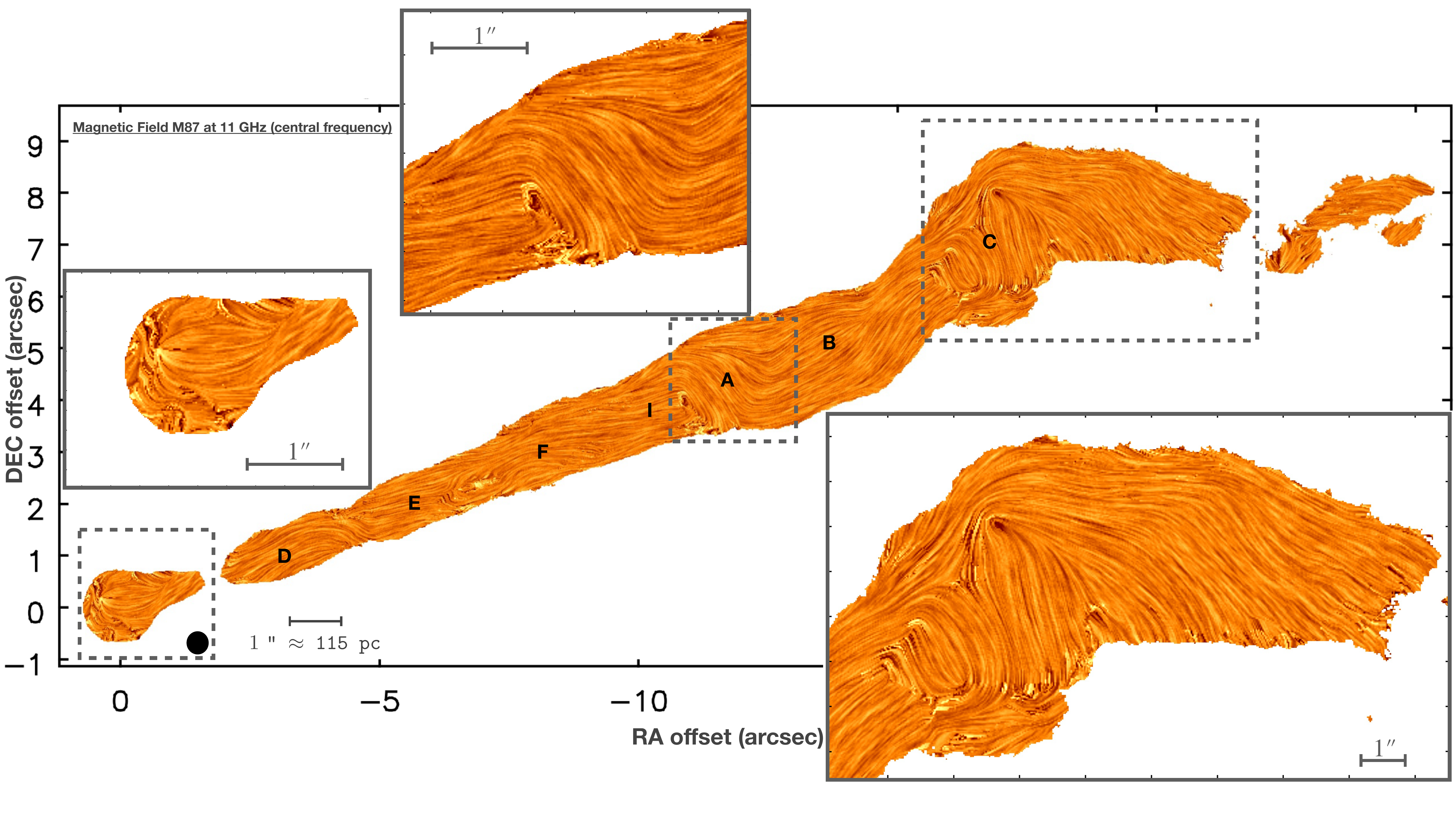}
\caption{Line integral convolution (LIC) map of the sky-projected magnetic field orientation of M87 and zooms of the core, knot A and knot C.}
\label{fig:BfieldZooms}
\end{figure}
\vspace{-6pt}
\begin{figure}[H]
\includegraphics[width=0.9\linewidth]{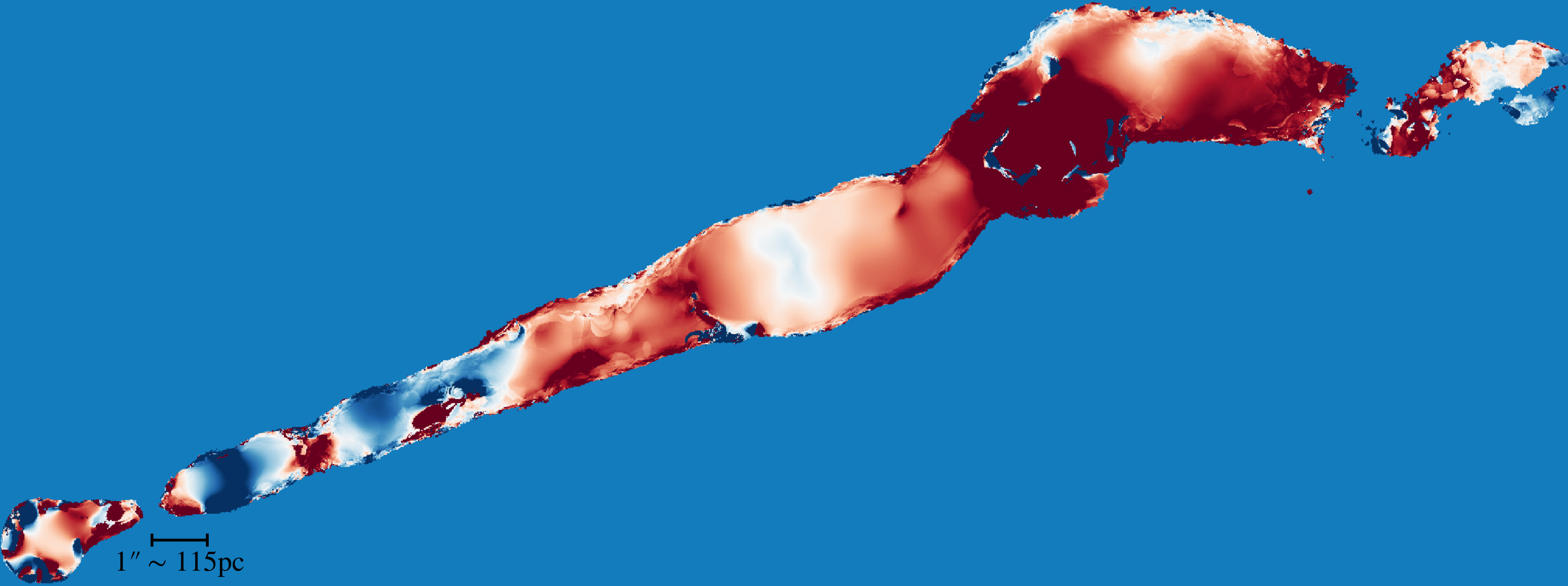}
\caption{ RM map of the kpc jet of M87. Color scale: from $-$500 to 500 rad/m$^2$. Variations on the RM values are detected.}
\label{fig:RMmap}
\end{figure}\vspace{-6pt}

\begin{figure}[H]
\includegraphics[width=0.7\textwidth]{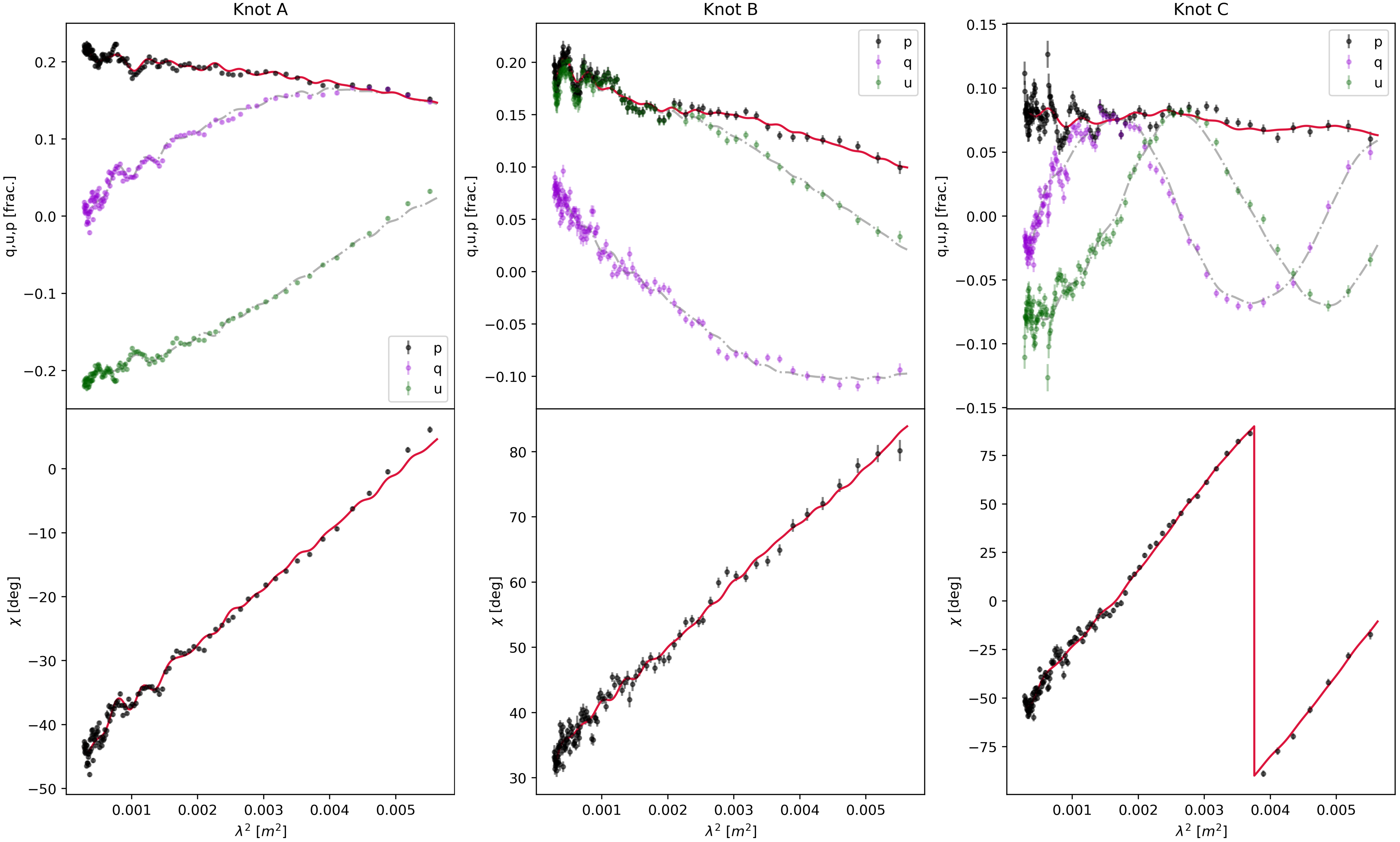}
\caption{Depolarization modelling of the knots A, B and C. Polarized Stokes parameters \textit{q} and \textit{u} of the three knots have been fitted using three internal Faraday screens. Top panels: the magenta points are the Stokes \textit{q}, the green points are the Stokes \textit{u} and the black points are the fractional polarization \textit{p}. Bottom panels: the black points are the polarized angle $\chi$ [deg]. Gray dot-dashed lines and red straight lines are the resulted model fit.}
\label{fig:DepolABC}
\end{figure}\vspace{-6pt}


\end{paracol}
\nointerlineskip
\begin{specialtable}[H]
\tablesize{\small}
 \widetable
\caption{Depolarization parameters resulted from the model fit. Integration has been performed over the knots}
\label{tab:DepolABC}
\setlength{\cellWidtha}{\columnwidth/14-2\tabcolsep-0in}
\setlength{\cellWidthb}{\columnwidth/14-2\tabcolsep-0in}
\setlength{\cellWidthc}{\columnwidth/14-2\tabcolsep-0in}
\setlength{\cellWidthd}{\columnwidth/14-2\tabcolsep-0in}
\setlength{\cellWidthe}{\columnwidth/14-2\tabcolsep-0in}
\setlength{\cellWidthf}{\columnwidth/14-2\tabcolsep-0in}
\setlength{\cellWidthg}{\columnwidth/14-2\tabcolsep-0in}
\setlength{\cellWidthh}{\columnwidth/14-2\tabcolsep-0in}
\setlength{\cellWidthi}{\columnwidth/14-2\tabcolsep-0in}
\setlength{\cellWidthj}{\columnwidth/14-2\tabcolsep-0in}
\setlength{\cellWidthk}{\columnwidth/14-2\tabcolsep-0in}
\setlength{\cellWidthl}{\columnwidth/14-2\tabcolsep-0in}
\setlength{\cellWidthm}{\columnwidth/14-2\tabcolsep-0in}
\setlength{\cellWidthn}{\columnwidth/14-2\tabcolsep-0in}
\scalebox{1}[1]{\begin{tabularx}{\columnwidth}{>{\PreserveBackslash\centering}m{\cellWidtha}>{\PreserveBackslash\centering}m{\cellWidthb}>{\PreserveBackslash\centering}m{\cellWidthc}>{\PreserveBackslash\centering}m{\cellWidthd}>{\PreserveBackslash\centering}m{\cellWidthe}>{\PreserveBackslash\centering}m{\cellWidthf}>{\PreserveBackslash\centering}m{\cellWidthg}>{\PreserveBackslash\centering}m{\cellWidthh}>{\PreserveBackslash\centering}m{\cellWidthi}>{\PreserveBackslash\centering}m{\cellWidthj}>{\PreserveBackslash\centering}m{\cellWidthk}>{\PreserveBackslash\centering}m{\cellWidthl}>{\PreserveBackslash\centering}m{\cellWidthm}>{\PreserveBackslash\centering}m{\cellWidthn}}
\toprule
\multirow{2}{*}{\vspace{-8pt}\textbf{Knot}} & \multirow{2}{*}{\vspace{-6pt}\textbf{p\boldmath$_{01}$}} & \multirow{2}{*}{\vspace{-6pt}\textbf{p\boldmath$_{02}$}} & \multirow{2}{*}{\vspace{-6pt}\textbf{p\boldmath$_{03}$}} &\textbf{\boldmath$\chi_{01}$} & \textbf{\boldmath$\chi_{02}$} & \textbf{\boldmath$\chi_{03}$} & \textbf{RM\boldmath$_{01}$} & \textbf{RM\boldmath$_{02}$} & \textbf{RM\boldmath$_{03}$} & \textbf{\boldmath$\Delta$RM\boldmath$_{01}$} & \textbf{\boldmath$\Delta$RM\boldmath$_{02}$} & \textbf{\boldmath$\Delta$RM\boldmath$_{03}$} & \multirow{2}{*}{\vspace{-8pt}\textbf{Std}} \\
\cmidrule{5-13}
     &          &          &          & \multicolumn{3}{c}{\textbf{[deg]}} & \multicolumn{6}{c}{\textbf{[rad/m\boldmath$^2$]}} & \\
\midrule
A & 0.20 & 0.11 & 0.11 & $-$45 & 49 & 34 & 155 & $-$1420 & $-$573 & $-$240 & 10,420 & 17,160 & 0.005 \\
B & 0.17 & 0.03 & 0.09 & 32 & 21 & $-$133 & 160 & 355 & $-$850 & $-$300 & 2270 & $-$17,460 & 0.007 \\
C & 0.01 & 0.08 & 0.06 & 120 & 113 & $-$0.1 & 1945 & 730 & $-$520 & $-$300 & $-$210 & 12,200 & 0.01 \\
\bottomrule

\end{tabularx}}
\end{specialtable}
\vspace{-13pt}
\begin{paracol}{2}
\switchcolumn

\subsection{Future Spectropolarimetric Surveys}

It is worth  mentioning three, already ongoing, important surveys which aim to inspect the polarized sky with the ultimate goal of investigating cosmic magnetism and the magnetic properties of individual sources: the Polarisation Sky Survey of the Universe's Magnetism (POSSUM) \citep{Gaensler2010},  the Q and U Observations at Cm wavelengths and Km baselines with the ATCA (QUOCKA) surveys and the VLA Sky Survey (VLASS).

The POSSUM survey is one of the eight major surveys to be undertaken by ASKAP. POSSUM will be a continuum polarization survey of the entire sky south of +30 degrees, covering the frequency range 1130--1430 MHz to an RMS sensitivity of 10 $\upmu$y/beam at 10-arcsec resolution. The main result will be a catalogue of Faraday RMs for $\sim$a million extragalactic radio sources. This dataset will help to determine the 3D geometry of the Milky Way's magnetic field, to test dynamo and other models for magnetic field generation, and to carry out a comprehensive census of magnetic fields as a function of redshift in galaxies, in clusters and in the overall intergalactic medium (for more information see: \url{https://askap.org/possum/}. 
The first results using commissioning data from POSSUM have been published by \cite{Anderson2021}, in which a first Faraday RM grid study of the Fornax cluster, which is presently undergoing a series of mergers, is presented. An RM grid is a powerful tool for probing foreground magnetic fields (i.e., cases where the emitting and Faraday rotating regions are well separated), and where the study of these external Faraday screens can be used to extract the global magneto-ionic properties of the intervening material \citep{Johnston-Hollitt2015}.

The QUOCKA project (\url{https://research.csiro.au/quocka/publications/}, accessed on August 2018), studies radio galaxies across the 1--8.5 GHz range, which is essential to uncovering their detailed polarized structure. The primary aims of the QUOCKA survey are to solve outstanding issues like understanding the complexity exhibited by polarised radio galaxies, and how the observed polarization properties are related to the internal structure and the evolution of the host objects. QUOCKA is building an extensive library of broadband polarization spectra, unveiling the internal structure of the sources, and revealing the connection to other source properties (e.g., environment, degree of activity, redshift, projected proximity to foreground sources). QUOCKA project progress is shared on the website: \url{https://research.csiro.au/quocka}.

In the northern sky, the most important ongoing survey is the VLA Sky Survey (VLASS), one of the largest all-sky radio observations in 40 years. It will map 80 percent of the sky in three phases over 7 years and is expected to catalog approximately 10 million radio sources in the frequency range of 2 to 4 GHz (S band). Its polarization data will be extremely useful for spectropolarimetric analysis of all the radio sources the survey will detect. For more information see {\url{https://science.nrao.edu/vlass/}}. 

Future surveys with Square Kilometre Array (SKA) are also planned (for a complete summary on the projects see: \cite{Gaensler2015}). For example, Beck et al. \cite{Beck2004} proposed a 1.4-GHz polarisation survey with the SKA covering 10,000 deg$^2$, resulting in a closely spaced Faraday RM grid of AGN.

\section{Summary and Conclusions}
\label{CONCLUSION}
I reviewed a technique for the analysis of broad-band spectropolarimetric data: Stokes QU-fitting. This technique consists of the modeling of both the Stokes parameters \textit{Q} and \textit{U} over a large range of wavelengths using analytical models available in literature. An overview of the polarization models has been presented. The principal characteristic is whether the change of the polarized plane (the Faraday depth) and the decreased polarized signal (the depolarization effects), is considered internal or external to the source of study. The presented models consider the presence of a regular or turbulent magnetic field configuration or a combination of both. The main characteristic of all the observational studies available over the last 10 years is the presence of complex behaviors of the polarized light; therefore, with an abrupt decrease of the fractional polarization with wavelength and a non linear behavior of the polarized angle with the wavelength. This is mainly due to multiple Faraday screens close to the sources of study (depending on the angular resolution of the observations) in which the magnetic field could be either regular or with a mixture of regular and turbulent magnetic fields. Radio sources are revealing themselves in their complexity also under the polarized light. 
The Stokes QU-fitting is an exquisite tool to analyze those complex substructures. A new messenger is available to decipher and to indirectly map the magneto--ionic medium of AGN.   

\vspace{6pt}


\funding{This research received no external funding.}

\institutionalreview{Not applicable}

\informedconsent{Not applicable}

\dataavailability{Not applicable}

\acknowledgments{ A.P. acknowledges the program C\'atedras CONACyT. }

\conflictsofinterest{The author declares no conflict of interest.} 
\end{paracol}
\reftitle{References}

\externalbibliography{yes}

\end{document}